\documentclass[prl,aps,twocolumn,showpacs,amssymb]{revtex4}

\usepackage{amsmath}

\def\idty{{\openone}}
\def\abs #1{\vert#1\vert}
\def\tr #1{\mathop{\rm tr}\left[#1\right]}

\def\norm #1{\Vert #1\Vert}

\def\Null{{\bf 0}}

\def\ketbra #1#2{\vert #1\rangle \negmedspace \langle #2\vert}

\begin{document}

\title{Hiding classical data in multi-partite quantum states}

\author{T.~Eggeling}
  \email{t.eggeling@tu-bs.de}
\author{R.~F. Werner}
  \email{r.werner@tu-bs.de}
\affiliation{Institut f{\"u}r Mathematische Physik, TU
Braunschweig,\\ Mendelssohnstr.3, 38106 Braunschweig, Germany.}

\begin{abstract}
We present a  general technique for hiding a classical bit in
multi-partite quantum states. The hidden bit, encoded in the
choice of one of two possible density operators, cannot be
recovered by local operations and classical communication without
quantum communication. The scheme remains secure if quantum
communication is allowed between certain partners, and can be
designed for any choice of quantum communication patterns to be
secure, but to allow near perfect recovery for all other patterns.
The maximal probability of unwanted recovery of the hidden bit, as
well as the maximal error for allowed recovery operations can be
chosen to be arbitrarily small, given sufficiently high
dimensional systems at each site. No  entanglement is needed since
the hiding states can be chosen to be separable. A single ebit of
prior entanglement is not sufficient to break the scheme.
\end{abstract}


\maketitle

\section{Introduction}
Many secrets in the world are locked away with keys distributed
among many parties. A well-known classical scheme for this is
Shamir's secret sharing \cite{Sh}, in which a pre-assigned
fraction of the key-possessing parties needs to contribute their
parts of the key to unlock the secrets. There are two directions
in which this can be generalized to hiding information in
multi-partite quantum states. In the first version called
``quantum secret sharing'', the bit is hidden in a way that some
parties can recover the bit via local operations and classical
communication \cite{QS1,QS2}. Typically the information is then
hidden in pure states and the theory is closely related to the
theory of error correcting codes, the errors corresponding to the
parties whose part of the key is not available. In the second
version, which has been called ``quantum data hiding''
\cite{DiV,DiV2}, and which we follow in this paper, one still
hides a classical bit, but the quantum structure is used to
increase the demands on the communication needed for the recovery.
Arbitrary classical communication between $N$ parties (along with
arbitrary local quantum operations) is allowed, but only with a
pre-assigned amount of quantum information exchange the hidden
information can be retrieved.

In \cite{DiV,DiV2} only the case $N=2$ was considered.
Since the hiding states have very high symmetry in that 
case (they are special ``Werner states'' \cite{W89}) DiVincenzo et
al.\ suggest that multi-partite (i.e., $N>2$) data hiding
scenarios might be based on highly symmetric multi-partite
entangled states such as the ones studied in our paper \cite{EW}.

Building on this idea we will generalize data hiding to an
extremely versatile scheme: For $N$-partite systems one can
freely choose for which patterns of quantum communication the
hidden bit can be retrieved, and for which patterns it remains
hidden. The level of security  can be chosen arbitrarily high: the
maximal probability of unwanted recoveries and probability for
erroneous identification using an allowed pattern of quantum
communication go to zero like the inverse of the dimension of the
Hilbert spaces at each site. Expressed in terms of the number of
hiding qubits this is exponentially good.

Surprisingly, no entanglement is needed for this scheme: the
hiding states can be chosen to be separable (this was strongly
suggested, but not proved in \cite{DiV,DiV2}). In keeping with
this, the scheme cannot be broken with a finite amount of prior
entanglement. For an entanglement based scheme one would expect
that hiding a single bit between two parties becomes insecure if
one ebit of prior entanglement is available to them. However, we
will show that the amount of entanglement needed to break security
is instead of the order needed to establish quantum communication
by teleportation.

In this letter we will focus on the main points of the
construction and the main ideas of the proof. For brevity, we will
give details only for the case of $N=4$ equivalent parties. Full
proofs of the case of general $N$ and parties possibly playing
different roles, will appear elsewhere \cite{future}.

\section{Main result}

Throughout we will assume that one classical bit has been encoded
in the preparation of a multi-partite quantum system, by preparing
either a density operator $\rho_0$ or $\rho_1$ on the Hilbert
space ${\cal H}_1\otimes\cdots\otimes{\cal H}_N$. We imagine the
$N$ subsystems to be distributed to widely separated laboratories.
The aim of the parties is to find out the value of the hidden bit.
For this they are allowed arbitrary classical communication and
can perform local quantum operations. In addition they may have
established quantum communication lines between some of the labs,
and their success will depend crucially on which quantum lines are
available. Since we do not distinguish between good and bad
quantum lines, this pattern of allowed quantum communication is
encoded in a {\it partition} ${\mathcal P}$  of the $N$ sites into
disjoint subsets: inside each of the subsets arbitrary quantum
communication is allowed, so these sites act like one party, but
no quantum communication is possible between sites in different
subsets of ${\mathcal P}$. For example, the partition ${\mathcal
P}=(\{1,2\},\{3,4\})$ means that sites $1$ and $2$ can exchange
quantum information freely, just like $3$ and $4$, but between
these groups only classical communication is allowed.

Whatever procedure the $N$ parties apply will amount to measuring
some ``analyzing operator'' $A$, $\Null\leq A\leq\openone$ such
that the probability for guessing the value ``1'' of the hidden
bit on an initial preparation $\rho$ is $\tr{\rho A}$. The
locality conditions imply that only certain operators $A$ are {\it
admissible for} ${\mathcal P}$. Of course, the parties will try to
make $\tr{\rho_1 A}\approx1$ and $\tr{\rho_0 A}\approx0.$ We say
that for a particular pair of hiding states $\rho_1,\rho_0$ a
partition ${\mathcal P}$ is {\it hiding with quality
$\varepsilon_1$}, if $\abs{\tr{(\rho_1-\rho_0)A}}\leq\varepsilon_1$
for all admissible analyzing operators $A$. On the other hand, we
say that ${\mathcal P}$ is {\it revealing with quality
$\varepsilon_2$}, if for some admissible $A$ we have
$\abs{\tr{(\rho_1-\rho_0)A}}\geq1-\varepsilon_2$.

Whoever is hiding the information does not know in advance what
communication pattern will be established. But, as our
construction will show, the states $\rho_1,\rho_0$ can be designed
such that, for any choice of $\varepsilon_1,\varepsilon_2>0$, every partition is
either hiding or revealing with quality $\varepsilon_1$ or $\varepsilon_2$ respectively.
The set of hiding partitions can be chosen arbitrarily subject {\it only} to
the trivial constraint that for every partition which is finer
than a hiding one, i.e., which corresponds to a pattern allowing
less quantum communication, must itself be hiding. We remark that
the Hilbert space dimensions need to become large if the $\varepsilon_i$
are small. In fact, in our construction the $\varepsilon_i$ typically
behave like $1/d$, if $d$ is the dimension of the one-site Hilbert
spaces. The construction naturally also yields {\it separable}
states $\rho_1,\rho_0$ satisfying the conditions, although for
these still higher dimensions $d$ are required to achieve the same
errors.

In this letter we will explicitly construct hiding states
$\rho_1,\rho_0$ for all choices of hiding partitions of $4$
parties, which are democratic in the sense that each site plays
the same role. It is remarkable that two such choices are not
comparable in the sense that neither allows more communication
than the other: we will give states for which any 2:2 partition
$(\{1,2\},\{3,4\})$ is hiding and any 3:1 partition
$(\{1,2,3\},\{4\})$ is revealing, but also states for which the
opposite is true.  Hence ``hiding strength'' of pairs of states
cannot be parametrized by a one-dimensional scale.

\section{Construction}
\subsection{Symmetric states}
We begin by restricting ourselves to a class of highly symmetric
states known as multipartite Werner states \cite{EW}. Their main
virtue is that they can be described by a fixed set of parameters
while the local Hilbert space dimensions go to infinity. By
definition, $4$-partite Werner states live on $({\mathbb C}^d
)^{\otimes 4})$, and  commute with all unitary operators of the
form $U^{\otimes 4}$ with $U$ a unitary operator on the
$d$-dimensional Hilbert space ${\mathbb C}^d$. This is equivalent
to the possibility of writing the state as linear combinations of
permutation operators (see \cite{Weyl}). For any permutation $\pi$
of the four sites we will denote the corresponding permutation
operator by $V_\pi:=\sum_{i,j,k,l=1}^{d}\ketbra{\pi(ijkl)}{ijkl}$.

Since the communication patterns we consider are invariant under
permutations we can even choose the states to be permutation
symmetric. We denote by $(i_1,i_2,\dots,i_r)$ the cyclic
permutation $i_1\mapsto i_2\mapsto \cdots\mapsto i_r\mapsto i_1$.
Then we must have, e.g.,
$\tr{\rho_iV_{(12)}}=\tr{\rho_iV_{(23)}}$, since these
permutations differ only by a relabelling of the sites. This
leaves just 4 expectations characterizing the state, namely
\begin{eqnarray}\label{permpars}
   r_2=\tr{\rho_iV_{(12)}}&\qquad&
   r_{22}=\tr{\rho_iV_{(12)(34)}}\nonumber\\
   r_3=\tr{\rho_iV_{(123)}}&\qquad&
   r_{4}=\tr{\rho_iV_{(1234)}}\;.
\end{eqnarray}
We will fix this vector $\vec\rho =(r_2,r_{22},r_3,r_4)$ of
expectations independently of the dimension $d$. Thus we
automatically get hiding schemes, which work for all dimensions,
though achieving $\varepsilon_1\to0$ only in the limit $d\to\infty$.
Whether or not a particular vector of expectations corresponds to
a family of density operators can be decided independently of
the dimension by group theoretical criteria, the extremal
possibilities being given by irreducible representations of the
permutation group. For details we refer to \cite{future}.

\subsection{Analyzing operators for fixed {}${\mathcal P}$}
Without loss of discriminating power we can then suppose that the
analyzing operators $A$ also have the $U^{\otimes 4}$ symmetry:
The $4$ parties only have to perform the same random unitary
rotation at every site (``twirling'') before realizing their
procedure. The resulting $A$ will commute with $U^{\otimes 4}$ but
will have exactly the same discriminating power for states
insensitive to such unitary rotations. Hence we can write
\begin{equation}\label{api}
  A=\sum_\pi a_\pi V_\pi
\end{equation}
 with suitable coefficients $a_\pi$. Note that this averaging
operation does {\it not} work for the permutation symmetry,
because the permutations are non-local operations, which would
clearly require the exchange of quantum information.

It turns out that in the sum (\ref{api}) we must distinguish two
types of terms depending on how the permutation $\pi$ relates to
the partition ${\mathcal P}$. We say that $\pi$ is {\it adapted}
to ${\mathcal P}$, if $\pi$ maps each of the sets in the partition
into itself. Clearly, if only the coefficients $a_\pi$ for $\pi$
adapted to ${\mathcal P}$ are non-zero, $A$ is a local operator in
this communication situation, hence admissible. Only such local
operators will be needed to show that certain patterns are
revealing in our theory.

The key problem (settled in the following subsection) is the
converse, namely to show that every operator $A$ which is
admissible for the partition ${\mathcal P}$ is at least
approximately of this sort. Fortunately, we can use here the same
simple criterion already employed in \cite{DiV,DiV2}, which is
based on {\it partial transposition}. The partial transpose
operation $\Theta_S$ associated with a subset
$S\subset\{1,2,3,4\}$ of the sites takes a tensor product operator
$A_1\otimes \cdots\otimes A_4$ to a similar product, in which all
$A_i$ with $i\in S$ are replaced by their matrix transpose in a
fixed basis. For example, $\Theta_{\{2,3\}}$ transposes only the
second and the third tensor factor of the input. The arguments in
\cite{DiV,DiV2} then tell us that, for any operator $A$, which is
admissible for ${\mathcal P}$, we must have that
\begin{equation}\label{pptcrit}
 \Null\leq\Theta_S(A)\leq\idty
\end{equation}
for all subsets $S$ compatible with $\mathcal{P},$ i.e., for all
$S$ which can be written as unions of the disjoint subsets forming
the partition $\mathcal{P}.$ Since positivity is preserved under
global transposition, it suffices to verify this for either $S$ or
its complement. For example, for
${\mathcal{P}}=(\{1\}\{2,3\}\{4\}),$ we must require
(\ref{pptcrit}) for the four subsets $S=$ (empty set), $\{1\},$
$\{2,3\},$ and $\{4\}.$

\subsection{Coefficients of admissible operators}
In this subsection we sketch the proof of the  following
Lemma:\newline
 {\it Suppose that $A$ is an analyzing operator, which is admissible for
the partition ${\mathcal{P}}$. Then in the sum (\ref{api}) all
coefficients $a_\pi$ with $\pi$ not adapted to ${\mathcal{P}}$ are
bounded by $c/d$, where $c$ is a constant depending only on $N.$ }

We will abbreviate by ${\mathbf O}(1/d)$ any terms bounded by a
constant times $1/d$, and leave the estimate of the constants to
\cite{future}. Consider the matrix $M$ given by $M_{\pi,\sigma}=
d^{-4}\tr{V^{*}_{\pi}V^{}_{\sigma}}.$ Then since $\tr{V_\pi}=d^c$,
where $c$ is the number of cycles in $\pi$ (including those of
length 1), we find $M_{\pi,\sigma}=\delta_{\pi,\sigma}+{\mathbf
O}(1/d)$. Thus to leading order in $d$, the permutation operators
are an orthonormal system with respect to the normalized trace.
Then by standard perturbation theory the matrix $M^{-1}$ is also
close to the identity, and we can approximately determine the
coefficients in the sum (\ref{api}) from
\begin{equation}\label{apiapprox}
    a_\pi=d^{-4}\tr{V^*_\pi A}+{\mathbf O}(1/d)\;.
\end{equation}

A crucial step in our estimate is to get the trace norm
($||X||_1=\tr{\sqrt{X^*X}}$) of
partially transposed permutation operators.
 We claim that
\begin{equation}\label{tS1}
    ||\Theta_S(V_{\pi})||_1=d^{4-l_S(\pi)} \;,
\end{equation}
where $l_S(\pi)$ denotes the number of points in $S$, which are
mapped outside $S.$ Rather than proving this in
general, consider as an example the case $S=\{1,2\}$ and
$\pi=(2,3)$. Since `1' is fixed and `2' is mapped to `3' outside
$S$, we have $l_S(\pi)=1$. We can write
$\Theta_S(V_{\pi})=\Theta_S(\sum_{ijnm}|ijnm\rangle\langle injm|)=
  \sum_{ijnm}|innm\rangle\langle ijjm|$. This can be written as
$d\; \openone\otimes P^{(23)}\otimes\openone$, where $P^{(23)}$
denotes the one dimensional projection onto the maximally
entangled vector on sites 2 and 3. Thus $\Theta_S(V_{\pi})$ has only
the non-zero eigenvalue $d$ with multiplicity $d^2$. This gives
$||\Theta_S(V_{\pi})||_1=d^3$ as claimed. More generally, $l_S(\pi)$
appears in this computation as the number of repeated indices in
either ket or bra in the analogous representation of
$\Theta_S(V_{\pi})$.

We now apply the standard estimate $\tr{XY}\leq||X||_1\cdot||Y||$,
and use that taking a partial transpose of both $X$ and $Y$ does
not change the trace. Hence, if $||\Theta_S(A)||\leq1$,
\begin{eqnarray}\label{estim}
d^{-4}\abs{\tr{AV_\sigma}}&=&d^{-4}\abs{\tr{\Theta_S(A)\Theta_S(V_\sigma)}}
  \nonumber\\
 &\leq& d^{-4}\norm{\Theta_S(V_\sigma)}_1\norm{\Theta_S(A)}\leq d^{-l_S(\sigma)}.
\end{eqnarray}

Coming back to the statement of the Lemma: let $\pi$ not be
adapted to $\mathcal{P}$. Then there is some set $S$ of the
partition, which is not mapped into itself by $\pi$. For this set
$l_S(\pi)\geq1$. On the other hand, since $A$ is admissible for
$\mathcal{P}$ the inequality (\ref{pptcrit}) must hold for this
$S$, hence $||\Theta_S(A)||\leq1$. Hence by combining
(\ref{apiapprox}) with (\ref{estim}) we get
$|a_\pi|
  =d^{-4}\abs{\tr{AV_\sigma}}+{\mathbf O}(1/d)
  \leq d^{-l_S(\pi)}+{\mathbf O}(1/d)
  ={\mathbf O}(1/d).$

\subsection{Tailoring the states}
 The idea of the construction is to choose $\rho_1$ and $\rho_0$ so
that $\tr{\rho_1V_\pi}=\tr{\rho_0V_\pi},$ for all permutations
$\pi$ which are adapted to {\it any} of the targeted hiding
partitions $\mathcal P$. Thus when we insert (\ref{api}) into
$\tr{(\rho_1-\rho_0)A}$ for any $A$ admissible for $\mathcal P$
the only contributing coefficient are $a_\pi={\mathbf O}(1/d)$.
Hence the whole expectation goes to zero.

On the other hand, we will make sure that
$\tr{(\rho_1-\rho_0)V_\pi}\neq0,$ for at least one permutation
adapted to each of the targeted revealing partitions. From this we
get an admissible analyzing operator with analyzing quality
$\varepsilon_2\neq0$, and independent of $d$. Analysis may not be
with probability one, but imperfect analysis can always be
upgraded to certainty as described in the following section.

\subsection{Verifying the examples}
 In the following examples the hiding states are given in terms
of the vector of expectations in (\ref{permpars}). The hiding
partitions in each example are the given partition, together with
all its permutations and all its refinements.

\noindent{\bf Weakest hiding}. The only permutation adapted to the
finest partition ${\mathcal P}=(\{1\},\{2\},\{3\},\{4\})$ is the
identity. Hence {\it any} way of fixing the expectations of
permutation operators gives a hiding pair of states. For example,
we can take $\rho_0$ (resp. $\rho_1$) as the normalized projection
to the Bose (=symmetric) subspace (resp.\ the   Fermi
(=antisymmetric) subspace) of $({\mathbb C}^d )^{\otimes 4}$. Thus
$\vec{\rho_0}=(1, 1, 1, 1)$ and $\vec{\rho_1}=(-1, 1, 1, -1).$
Obviously, if just two partners, e.g., 1 and 2, can exchange
quantum information they can find out which alternative $0/1$ was
chosen by just looking at the restriction of the state to their
pair of subsystems, and measuring ``symmetry''
$A=(\idty+V_{12})/2$.

\noindent{\bf Hiding against single pairs}.
 For all pair partitions ${\cal P}=(\{1,2\},\{3\},\{4\})$ the states
$\vec{\rho_0}=\frac{1}{3}(-1, -1, 0, 1)$ and
$\vec{\rho_1}=\frac{1}{3}(-1, 3, 0, -1)$ are hiding. Analysis for
`single pairs' and `triplets' (see below) is imperfect.

\noindent{\bf Hiding against two pairs}.
 For all partitions like ${\cal
P}=(\{1,2\},\{3,4\})$, the states $\vec{\rho_0}=(0, 1, 1, 0)$ and
$\vec{\rho_1}=(0, 1,-\frac{1}{2},0)$ are hiding. However, a
partition $(\{1,2,3\},\{4\})$ can use
$A=\frac{1}{3}(\idty+V_{(123)}+V_{(321)})$, to distinguish these
with certainty.

\noindent{\bf Hiding against triplets}.
 Conversely,  the states
$\vec{\rho_0}=\frac{1}{3}(3, 1, 0, 3)$ and
$\vec{\rho_1}=\frac{1}{3}(1, -1, 0, -1)$ are hiding for any
partition like ${\cal P}=(\{1,2,3\},\{4\})$, but can be analyzed
(imperfectly) by two pairs.

\noindent{\bf Strongest hiding} Finally, the states
$\vec{\rho_0}=\frac{1}{4}(0, 0, 1, 2)$ and
$\vec{\rho_1}=\frac{1}{4}(0, 0, 1, -2)$ are hiding unless quantum
communication between all parties is established, in which case
they can be distinguished perfectly.

\section{Multiple copies enhance recovery}
 As these examples show, our construction so far does not guarantee
perfect distinction ($\varepsilon_2=0$) for the partitions meant
to be revealing. However, there is a single device to boost the
detection quality, namely to distribute several, say $K$ copies of
the $N$-particle system, all prepared in the same state. Then for
the hiding partitions we still get $\varepsilon_1={\mathbf
O}(1/d).$ On the other hand, for the revealing partitions we can
use detection operators $A$ which are linear combinations of
permutations. Then the detection probabilities $\tr{\rho_1A}$ and
$\tr{\rho_0A}$ are independent of $d$, and if they are at all
different, measuring $A$ on all $K$ copies distinguishes $\rho_1$
and $\rho_0$ with any desired degree of certainty.

This shows that for getting good discrimination
$\varepsilon_2\to1$ we do not really need orthogonal states. What
counts is that $\rho_0$ and $\rho_1$ are different along
appropriate directions. Thus they can even be chosen to be close
to the maximally mixed state and, in particular, separable. Since this was
conjectured in \cite{DiV,DiV2} we include an explicit example,
namely the bipartite ($N=2$) case of our construction. At the same
time this illustrates nicely the interplay between the parameters
$d$ and $K$.

We use a simplified (but slightly weaker) bound to establish
hiding: Since all admissible analyzing operators satisfy
$\Null\leq \Theta_{\{2\}}(A)\leq\idty$, we get
 $\abs{\tr{(\rho_1-\rho_0)A}}
  = \abs{\tr{ \Theta_{\{2\}}(\rho_1-\rho_0) \Theta_{\{2\}}(A)}}
  \leq\norm{\Theta_{\{2\}}(\rho_1-\rho_0)}_1.$

Our single copy scheme is based on bipartite Werner states. With
$P_\pm$ the anti/symmetric projectors on ${\mathbb
C}^d\otimes{\mathbb C}^d$  and $\rho_\pm=P_\pm/\tr{P_\pm}$ our
hiding states are:
\begin{equation}
 \widehat\rho_0=\left(\frac{\rho_++\rho_-}2\right)^{\otimes K},\qquad
 \widehat\rho_1=\rho_+^{\otimes K},
\end{equation}
 which are clearly separable \cite{W89}.
From this one can readily compute the partial transposes
$\Theta_{\{2\}}(\rho_i)^{\otimes K}$ and their trace norm
difference, as well as the expectations of the analyzing operator
$A=P_+^{\otimes K}$, to get:
 \begin{equation}\label{biparted}
  \varepsilon_1=1-(1-1/d)^K
  ,\qquad {\rm and}\quad
  \varepsilon_2=2^{-K}.
\end{equation}
Thus we can first choose $K$ large to make $\varepsilon_2$ small, and
subsequently $d$ large, to get $\varepsilon_1=K/d+{\bf O}(d^{-2})$
small.

This separable scheme is remarkably robust even if the analyzing
partners share some entanglement:  If they share a  maximally
entangled pair of a $D$-dimensional system with fixed $D$, we get
the same asymptotic behaviour in the limit $d\to\infty$, just with
worse constants. Only if we choose $D$ to grow on the same scale
as $d$, i.e., on the same scale which would make teleportation
possible, we find that hiding becomes impossible.

Funding by the European Union project EQUIP (contract
IST-1999-11053) and financial support from the DFG (Bonn) is
gratefully acknowledged.

\end{document}